\documentclass[10pt,aps,prl,twocolumn,amsmath,amssymb,superscriptaddress,showpacs]{revtex4-1}

\usepackage{graphicx}
\usepackage{bm}
\usepackage[utf8]{inputenc}
\usepackage[T1]{fontenc}
\usepackage[english]{babel}
\usepackage{tikz}
\usepackage{pdfpages}

\begin{document}

\title{Two-dimensional analysis of the double-resonant $2D$ Raman mode in bilayer graphene}

\author{Felix Herziger}
\email{fhz@physik.tu-berlin.de}
\affiliation{Institut f\"ur Festk\"orperphysik, Technische Universit\"at Berlin, Hardenbergstrasse 36, 10623 Berlin, Germany}

\author{Matteo Calandra}
\affiliation{Institut de Min\'eralogie, de Physique des Mat\'eriaux, et de Cosmochimie, UMR CNRS 7590, Sorbonne Universit\'es - UPMC Univ Paris 06, MNHN, IRD, 4 Place Jussieu, F-75005 Paris, France}

\author{Paola Gava}
\affiliation{Institut de Min\'eralogie, de Physique des Mat\'eriaux, et de Cosmochimie, UMR CNRS 7590, Sorbonne Universit\'es - UPMC Univ Paris 06, MNHN, IRD, 4 Place Jussieu, F-75005 Paris, France}

\author{Patrick May}
\affiliation{Institut f\"ur Festk\"orperphysik, Technische Universit\"at Berlin, Hardenbergstrasse 36, 10623 Berlin, Germany}

\author{Michele Lazzeri}
\affiliation{Institut de Min\'eralogie, de Physique des Mat\'eriaux, et de Cosmochimie, UMR CNRS 7590, Sorbonne Universit\'es - UPMC Univ Paris 06, MNHN, IRD, 4 Place Jussieu, F-75005 Paris, France}

\author{Francesco Mauri}
\affiliation{Institut de Min\'eralogie, de Physique des Mat\'eriaux, et de Cosmochimie, UMR CNRS 7590, Sorbonne Universit\'es - UPMC Univ Paris 06, MNHN, IRD, 4 Place Jussieu, F-75005 Paris, France}

\author{Janina Maultzsch}
\affiliation{Institut f\"ur Festk\"orperphysik, Technische Universit\"at Berlin, Hardenbergstrasse 36, 10623 Berlin, Germany}


\begin{abstract}
By computing the double-resonant Raman scattering cross-section completely from first principles and including electron-electron interaction at the $GW$ level, we unravel the dominant contributions for the double-resonant $2D$-mode in bilayer graphene. We show that, in contrast to previous works, the so-called inner processes are dominant and that the $2D$-mode lineshape is described by three dominant resonances around the $K$ point. We show that the splitting of the TO phonon branch in $\Gamma-K$ direction, as large as 12\,cm$^{-1}$ in $GW$ approximation, is of great importance for a thorough description of the $2D$-mode lineshape. Finally, we present a method to extract the TO phonon splitting and the splitting of the electronic bands from experimental data.
\end{abstract}

\maketitle

Double-resonance Raman spectroscopy provides a versatile tool for investigating the electronic structure and phonon dispersion of graphitic systems by tuning the laser energy \cite{PhysRevLett.85.5214}. In particular, the $D$ and $2D$ Raman modes allow to investigate structural changes, such as the number of layers, disorder, strain and doping in the sample \cite{PhysRevLett.97.187401, 10.1021/nl8031444, PhysRevB.79.205433, PhysRevB.82.201409, 10.1073/pnas.0811754106, 10.1021/nl102123c, 10.1021/nn103493g}. 

Especially the distinction between single, bi-, and few-layer graphene via measuring the $2D$ mode attracted great attention due to its simplicity \cite{PhysRevLett.97.187401}. In single-layer graphene, the double-resonance is often simplified to one single scattering process, well describing the experimental peak shape of the $2D$ mode. Up to now, the $2D$ mode in bilayer graphene is described and interpreted within the framework of four scattering processes. Each process was assigned to a different spectral feature in the $2D$-mode lineshape, phenomenologically explaining the observed peakshape \cite{PhysRevLett.97.187401}. All successive studies on the $2D$ mode in bilayer graphene relied on this assignment \cite{PhysRevB.76.201401,PhysRevB.77.245408, PhysRevB.80.241414, 10.1016/j.carbon.2010.11.053, 10.1021/nl300477n,10.1002/jrs.2435}. Furthermore, the $2D$ mode in bilayer graphene has been mainly discussed in terms of outer processes \cite{PhysRevLett.97.187401, PhysRevB.76.201401, PhysRevB.77.245408, PhysRevB.80.241414,10.1002/jrs.2435}. However, the importance of inner processes was shown both theoretically and experimentally for the $2D$ mode in single-layer graphene \cite{PhysRevB.82.201409,PhysRevB.84.035433,10.1021/nl400917e}. In bilayer graphene, only very few works considered the possibility of contributions from inner processes, but were still neglecting the splitting of the two TO (transversal optical) phonon branches \cite{10.1016/j.carbon.2010.11.053, 10.1021/nl300477n}. Hence, the role of different contributions to the double-resonance in bilayer graphene is still under discussion and needs final clarification. 

In this letter, by completely calculating the double-resonant Raman cross-section from first principles and by comparing with experimental spectra for different laser energies, we unravel the dominant scattering processes in bilayer graphene. In contrast to previous works that explained the $2D$-mode lineshape with four independent scattering processes \cite{PhysRevLett.97.187401, PhysRevB.76.201401,PhysRevB.77.245408, PhysRevB.80.241414, 10.1016/j.carbon.2010.11.053, 10.1021/nl300477n,10.1002/jrs.2435}, we show that the $2D$ mode is described by three dominant resonances around the $K$ point from inner processes plus a weaker contribution from outer processes. We show that the $GW$ correction to the TO phonon branch leads to a much larger TO splitting than in LDA approximation. This splitting cannot be neglected; we present an analysis to directly derive the TO phonon and electronic splitting in bilayer graphene with high accuracy.


Experimental Raman spectra were obtained from freestanding bilayer graphene in back-scattering geometry under ambient conditions using a Horiba HR800 spectrometer with a 1800\,lines/mm grating with spectral resolution of 1\,cm$^{-1}$. During all measurements the laser power was kept below 0.5\,mW to avoid sample damaging or heating. Spectra were calibrated by standard neon lines. The freestanding bilayer graphene enables us to probe the intrinsic $2D$-mode lineshape, ensuring an accurate extraction of the fitting parameters \cite{10.1021/nl400917e}, following the model of Basko \cite{PhysRevB.78.125418}.

\begin{figure*}%
\includegraphics[width=\textwidth]{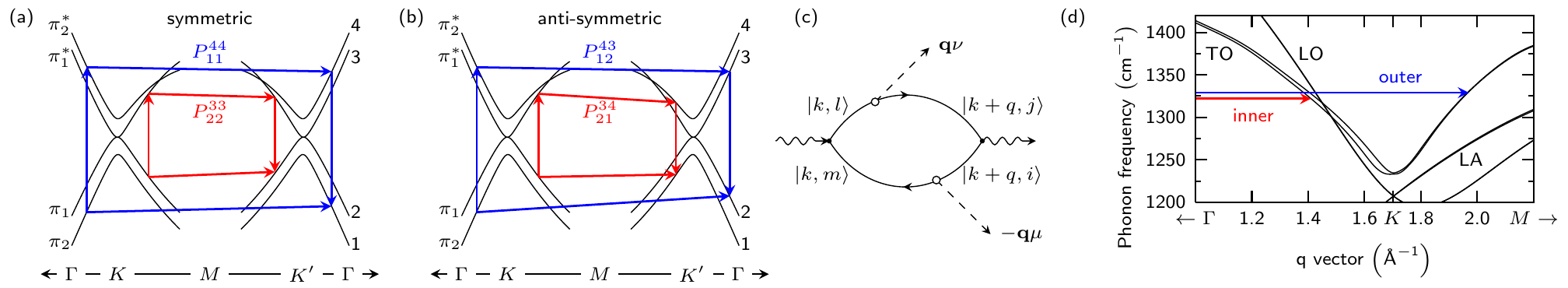}%
\caption{Schematized illustration of the $2D$-mode scattering processes along the $\Gamma-K-M-K'-\Gamma$ high-symmetry direction for (a) symmetric and (b) anti-symmetric processes. Inner and outer processes are marked with red and blue traces, respectively. (c) Goldstone diagram for a double-resonant \textit{e-h} scattering process $P_{mi}^{lj}$. (d) $GW$-corrected phonon dispersion of bilayer graphene close to $K$, showing the TO splitting in $\Gamma-K$ direction.}%
\label{fig:DRsketch}%
\end{figure*}


In Bernal-stacked bilayer graphene the $\pi$ orbitals give rise to two valence and two conduction bands, denoted as $\pi_{1}$, $\pi_{2}$ and $\pi^{*}_{1}$, $\pi^{*}_{2}$. Bilayer graphene possesses two TO phonon branches, each one degenerate with an LO (longitudinal optical) branch at $\Gamma$. At $\Gamma$ the TO branches are split into a symmetric and an anti-symmetric vibration. The symmetric TO phonon is an in-phase vibration between the lower and upper layer and exhibits $E_g$ symmetry in the point group $D_{3d}$, whereas the anti-symmetric vibration is out-of-phase ($E_u$ symmetry). Our calculated frequency splitting at $\Gamma$ is approx. 5\,cm$^{-1}$, comparable with the experimentally observed 6\,cm$^{-1}$ splitting in graphite \cite{10.1016/0008-6223(95)00035-C}. Along the $\Gamma-K$ direction, the $GW$-calculated TO splitting increases to values as large as 12\,cm$^{-1}$, whereas the splitting in LDA is approximately two times smaller. The displacement patterns of the TO vibrations change away from $\Gamma$, we however extend the labeling of the phonon branches throughout the BZ.

\begin{figure}%
\includegraphics[width=\columnwidth]{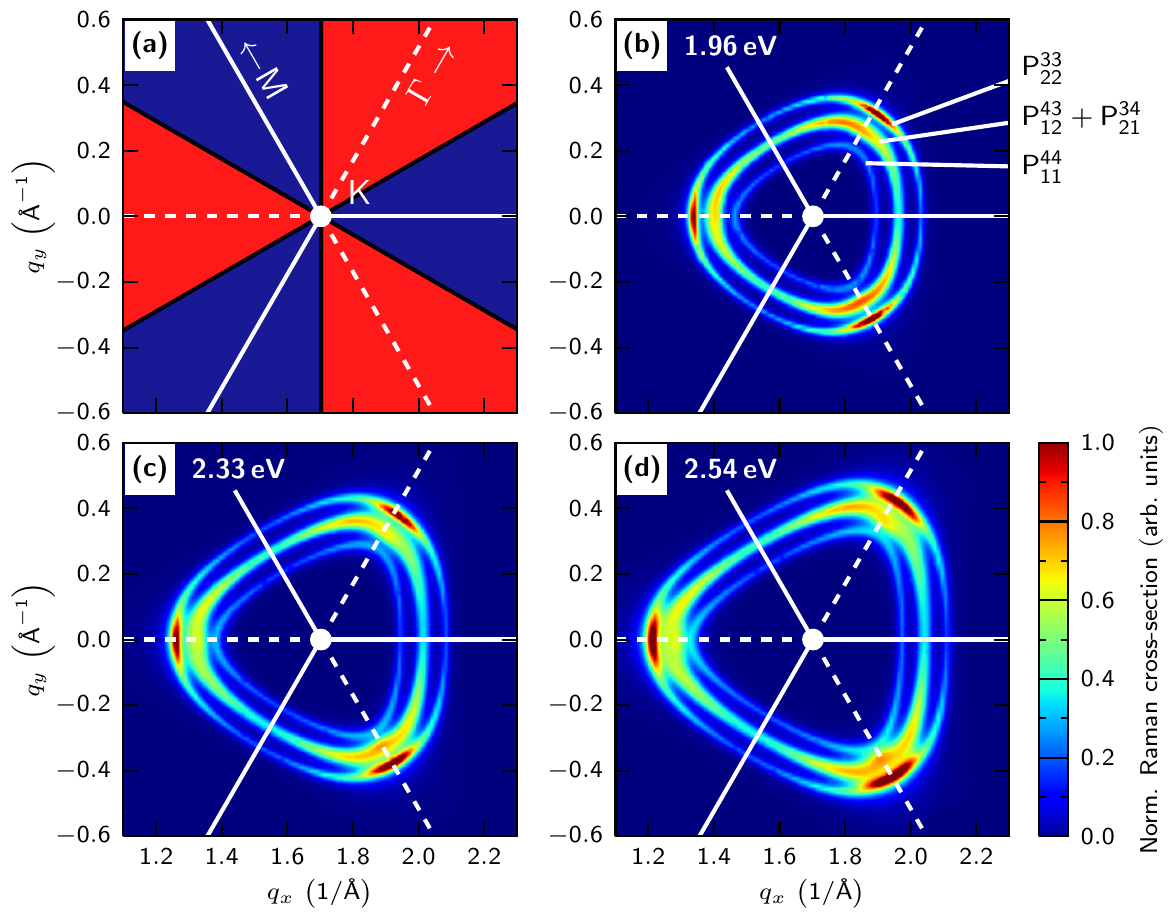}%
\caption{(a) Illustration of the phonon wave-vector sectors for inner (red) and outer (blue) processes around the \textit{K} point. The solid and dashed white lines denote the $K-M$ and $K-\Gamma$ high-symmetry lines, respectively. (b)-(d) Plots of the normalized $2D$-mode scattering cross-section ${\cal I}_{\bf q}$ around the $K$ point for different laser energies.}%
\label{fig:Contour}%
\end{figure}

The double-resonant $2D$ mode is a second-order Raman process, involving two TO phonons with wave vector $\mathbf{q}\neq0$. The process can be divided into four virtual transitions, (i) creation of an electron-hole pair by a photon with energy $\hbar\omega_L$, (ii) scattering of an electron/hole state by a phonon with wave vector $\mathbf{q}$, (iii) scattering of an electron/hole state by a phonon with wave vector $-\mathbf{q}$ and, (iv) recombination of the electron-hole pair. The observed frequency of this process is twice the phonon frequency at $\mathbf{q}$. As we have explicitly verified, the processes where one phonon is scattered by an electron and one phonon is scattered by a hole (diagrams of the kind shown in Fig. \ref{fig:DRsketch}), are, by far, the most dominant contribution to the Raman cross-section \cite{PhysRevB.84.035433}. We will refer to these processes as electron-hole scattering (\textit{e-h} scattering). The scattering processes can be further divided into symmetric/anti-symmetric and inner/outer processes. Symmetric processes are scattering events between equivalent electronic bands at $K$ and $K'$, whereas for an anti-symmetric process the band index is changing. We refer to the term inner process, if the resonant phonon wave-vector stems from a sector of $\pm 30^{\circ}$ next to the $K-\Gamma$ direction with respect to $K$. Conversely, outer processes have phonon wave-vectors from $\pm30^{\circ}$ next to the $K-M$ directions [Fig. \ref{fig:Contour}(a)]. To simplify the labeling of the scattering processes, we enumerate the electronic bands starting from the energetically lowest band near $K$. Every scattering process $P_{mi}^{lj}$ is then uniquely defined by four indices that are given by the band indices of the initial electron $m$, of the excited electron $l$, of the scattered electron $j$, and of the scattered hole $i$. Since the incoming light couples mostly to only two ($1\rightarrow4$ and $2\rightarrow3$) of the four possible optical transitions, four different combinations of \textit{e-h} scattering are allowed \cite{PhysRevLett.97.187401,PhysRevB.79.125426}. These are the symmetric processes $P_{11}^{44}$ and $P_{22}^{33}$ and the anti-symmetric processes $P_{12}^{43}$ and $P_{21}^{34}$.

Following Ref. \cite{PhysRevB.84.035433}, the two-phonon (\textit{pp}) double-resonant Raman intensity is
\begin{eqnarray}
I(\omega)&=&\frac{1}{N_q}\sum_{{\bf q},\nu,\mu}I^{pp}_{{\bf q}\nu\mu}
\delta(\omega_L-\omega-\omega_{\bf -q}^\nu-\omega_{\bf q}^\mu)\nonumber \\
&&[n(\omega_{\bf -q}^\nu)+1][n(\omega_{\bf q}^\mu)+1]
, \label{eq2} 
\end{eqnarray}
where $\omega_{\bf q}^\mu$ and $n(\omega_{\bf q}^\mu)$ are the phonon frequencies and the Bose distributions for mode $\mu$, respectively \footnote{We discuss here the scattering in terms of ``double-resonance'', since the dominant phonon momentum is determined by the two resonant conditions on the light absorption and emission (see, e.g., Sec. III E.2 and Fig. 23 of Ref. \cite{PhysRevB.84.035433}). However, triple-resonant processes also occur and are fully included in our calculations}. The probability of exciting two phonons is $I^{pp}_{{\bf q}\nu\mu} = \left|\frac{1}{N_{k}}\sum_{{\bf k},\beta}K^{pp}_\beta ({\bf k},{\bf q},\nu,\mu) \right|^2 $, where the matrix elements $K^{pp}_\beta ({\bf k},{\bf q},\nu,\mu)$ are defined by expressions involving the electron and phonon band dispersion, the electron-phonon coupling $g_{{\bf k}n,{\bf k}+{\bf q}m}^{\mu}$ and the electron-light $D_{{\bf k}n,{\bf k}m}$ matrix-elements throughout the full BZ (see appendix A of Ref. \cite{PhysRevB.84.035433}). Here, $\mathbf{k}$ refers to the electron wave-vector and $\beta$ labels the different possibilities of electron and hole scattering. We want to remind the reader of the importance of quantum interference in the double-resonance process. Scattering processes with the same final state $(\mathbf{q},\mu,\nu)$ but different intermediate states can observe interference. Consequently, scattering processes at different $\mathbf{q}$ do not interfere. In most previous works on the $2D$ mode in bilayer graphene, the interference between different processes was completely neglected. However, as will be shown later, quantum interference has remarkable impact on the $2D$-mode lineshape in bilayer graphene.

Due to the difficulties in obtaining $g_{{\bf k}n,{\bf k}+{\bf q}m}^{\mu}$ and $D_{{\bf k}n,{\bf k}m}$ directly from first principles, previous publications used matrix elements derived from tight-binding models \cite{PhysRevB.84.035433,10.1002/pssb.201100510,PhysRevB.85.115451}. Here, we overcome this difficulty by using Wannier interpolation \cite{10.1016/j.cpc.2007.11.016} of the electron-phonon and the electron-light matrix elements, as developed in Ref. \cite{PhysRevB.82.165111}. We first calculate from first principles in LDA approximation \cite{10.1088/0953-8984/21/39/395502} the unscreened electric dipole and the screened electron-phonon matrix elements on a $64\times 64$ electron-momentum grid and a $6\times6$ phonon momentum grid. We then interpolate them to denser $480\times 480$ electron-momentum grid randomly shifted from the origin and a $12288$-points phonon momentum grid, covering a sufficiently large region around the \textit{K} points. The phonon bands were Fourier interpolated from a $12\times 12$ phonon momentum grid. The electronic, the TO phonon bands, and $g_{{\bf k}n,{\bf k}+{\bf q}m}^{\mu}$ were $GW$-corrected, following the approach given in Ref. \cite{PhysRevB.84.035433} (see SI \cite{SI}). The electron broadening $\gamma$ was chosen to be twice as large as that in Ref. \cite{PhysRevB.84.035433} to account for additional electron-electron interaction \cite{PhysRevLett.112.207401}, namely $\gamma=0.081832\times(\hbar\omega_{L}/2 - 0.1645)$ eV. This choice gives better agreement with experiments (SI).  Finally, the $\delta$-function in Eq. \eqref{eq2} is broadened with an 8\,cm$^{-1}$ Lorentzian \cite{PhysRevB.87.214303}.

Fig. \ref{fig:Contour} presents calculated contour plots of ${\cal I}_{\bf q}=\sum_{\nu,\mu} I^{pp}_{{\bf q}\nu\mu}$ for the double-resonant $2D$ mode in bilayer graphene for various $\hbar \omega_L$. Three resonances around the $K$ point contribute to the $2D$ mode. These regions are attributed to, from inside to outside, the $P_{11}^{44}$, the anti-symmetric $P_{12}^{43}$ and $P_{21}^{34}$, and the $P_{22}^{33}$ processes. As the resonant phonon wave-vectors of the anti-symmetric processes are nearly degenerated, the resulting phonon frequencies are very similar, disproving previous assignments of anti-symmetric processes to different spectral features of the $2D$ mode \cite{PhysRevLett.97.187401, PhysRevB.76.201401,PhysRevB.77.245408, PhysRevB.80.241414, 10.1016/j.carbon.2010.11.053, 10.1021/nl300477n, 10.1002/jrs.2435}. Furthermore, the dominant contributions to the $2D$-mode scattering cross-section stem from the $K-\Gamma$ direction which can be identified with inner processes.

\begin{figure}%
\includegraphics[width=\columnwidth]{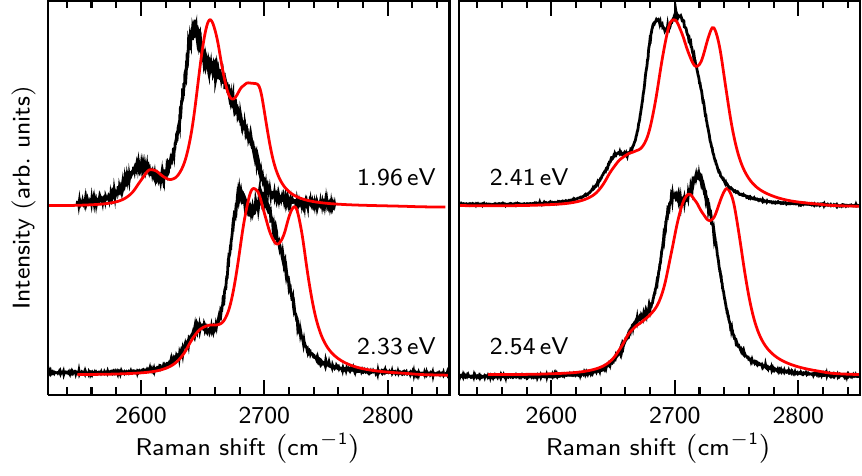}%
\caption{Comparison of calculated $2D$-mode spectra with Raman spectra from freestanding bilayer graphene at different $\hbar\omega_L$. Calculations and experimental data are shown as red and black curves, respectively. Spectra are normalized and vertically offset.}%
\label{fig:CalcExp}%
\end{figure}

\begin{figure*}%
\includegraphics{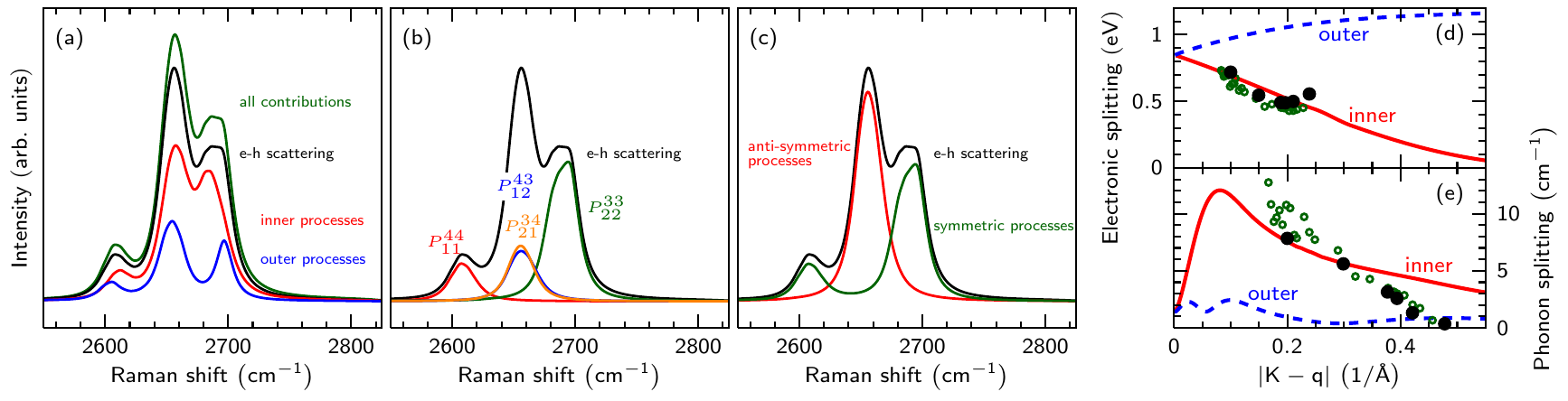}%
\caption{Calculated $2D$-mode spectra at 1.96\,eV excitation energy. Decomposition of the calculated spectra into (a) \textit{e-h} scattering processes, as well as inner and outer contributions, (b) the four different scattering processes $P_{mi}^{lj}$ (without interference between the different processes), and (c) into symmetric and anti-symmetric processes (including interference). (d) and (e) show the experimental values for the electronic and TO phonon splitting, respectively. The solid (red) and dashed (blue) lines denote the DFT calculated splittings in inner and outer direction, respectively. Open, green circles are data points from Ref. \cite{PhysRevB.80.241414}. Filled, black circles represent data points from this work.}%
\label{fig:DecompSplitting}%
\end{figure*}

We will now turn our discussion to the calculated Raman spectra of the $2D$ mode in bilayer graphene. Fig. \ref{fig:CalcExp} compares the calculated Raman spectra with spectra from freestanding bilayer graphene at different $\hbar\omega_L$. The overall agreement between calculation and experimental data is very good, although there is a slight mismatch in frequency. The calculated frequencies are approximately 10\,cm$^{-1}$ too high, yet our calculations reproduce the lineshape of the $2D$ mode, \textit{i.e.}, the relative intensities of the different contributions, very well. 

Fig. \ref{fig:DecompSplitting} shows the decomposition of the calculated $2D$-mode spectrum at 1.96\,eV excitation energy into its different contributions. The decomposition for other $\hbar\omega_L$ is accordingly (SI). As in single-layer graphene, we confirm that in bilayer graphene the \textit{e-h} scattering processes dominate compared to all other scattering paths. Furthermore, inner processes dominate over outer ones. By explicitly decomposing the $2D$ mode into the four different processes in Fig. \ref{fig:DecompSplitting}(b), we find that the symmetric $P_{11}^{44}$ and $P_{22}^{33}$ processes are on the low- and high-frequency side of the $2D$ mode, respectively. The frequencies of the anti-symmetric processes are in between the symmetric contributions and nearly degenerate, as already inferred from Fig. \ref{fig:Contour}. This disagrees with all previous works \cite{PhysRevLett.97.187401, PhysRevB.76.201401,PhysRevB.77.245408, PhysRevB.80.241414, 10.1016/j.carbon.2010.11.053, 10.1021/nl300477n, 10.1002/jrs.2435}, attributing substantially different phonon frequencies to the anti-symmetric processes. As seen in Fig. \ref{fig:DecompSplitting}(b), the decomposition of the single processes is not additive, \textit{i.e.}, the sum of the four processes does not yield the total spectrum. This can be directly attributed to quantum interference effects between the anti-symmetric processes. By decomposing the spectrum into the single processes as in Fig. \ref{fig:DecompSplitting}(b), interference between the $P_{mi}^{lj}$ is prohibited. However, $P_{12}^{43}$ and $P_{21}^{34}$ exhibit a large overlap in reciprocal space and interfere constructively. Decomposing the total spectrum into symmetric and anti-symmetric contributions and thus allowing interference between those processes yields the spectrum in Fig. \ref{fig:DecompSplitting}(c). This decomposition is additive. The constructive interference has remarkable impact on the $2D$-mode lineshape, \textit{i.e.}, the intensity of the anti-symmetric processes is drastically enhanced, highlighting the importance of quantum interference effects in the double-resonance process.

Up to now, we described the $2D$ mode in terms of three dominant resonances that split up into inner and outer contributions. Thus, one might expect six separate peaks in the $2D$-mode spectrum in total. This is in contrast to the experimentally observed lineshape, where usually three or four peaks can be distinguished. However, the decomposition in Fig. \ref{fig:DecompSplitting}(a) and (b) shows that inner and outer contributions for the $P_{11}^{44}$, $P_{12}^{43}$, and $P_{21}^{34}$ processes are nearly degenerate in frequency, thus reducing the number of observable $2D$-mode peaks for these processes to two. Only the $P_{22}^{33}$ process exhibits a splitting between inner and outer contributions that is large enough to be detected in experiments; it is responsible for the third and fourth peak in the $2D$-mode lineshape. Therefore, in experiments the $2D$ mode should be fitted with four peaks, where the assignment of the peaks, from lowest to highest frequency, is $P_{11}^{44}$, $P_{12}^{43}/P_{21}^{34}$, inner $P_{22}^{33}$, and outer $P_{22}^{33}$. In previous works, the inner $P_{22}^{33}$ contribution was erroneously assigned to an anti-symmetric scattering process, whereas the outer contribution, \textit{i.e.}, the small high-frequency shoulder of the $2D$ mode, was attributed to a symmetric process. Here, we showed that these two peaks result from the same scattering process ($P_{22}^{33}$). Our assignment of the third and fourth $2D$-mode peaks to inner and outer $P_{22}^{33}$ contributions is supported by recent experiments on strained bilayer graphene \cite{10.1021/nl203565p}. Due to different dispersions of inner and outer processes, both contributions to $P_{22}^{33}$ merge with increasing laser energy. Therefore, at higher laser energies the fourth peak vanishes. This can be seen in the spectrum of the freestanding bilayer graphene at 2.54\,eV excitation energy in Fig. \ref{fig:CalcExp}. Here, the small high-frequency shoulder cannot be identified any more, giving further evidence to our assignment of the three dominant contributions to the $2D$ mode in bilayer graphene.

In previous works, the TO splitting in bilayer graphene has always been neglected in the double-resonance, as only outer processes were considered and the TO splitting along $K-M$ is of the order of 1.5\,cm$^{-1}$ \cite{PhysRevLett.97.187401}. However, we proved that inner processes are dominant. In fact, along $K-\Gamma$ the TO splitting is as large as 12\,cm$^{-1}$ in $GW$ approximation. We observe that the dominant contributions to symmetric processes stem from scattering with symmetric TO phonons, whereas the dominant contributions to anti-symmetric processes result from scattering with anti-symmetric TO phonons (see SI).

The fact that symmetric and anti-symmetric processes couple to different phonon branches has remarkable impact on the $2D$-mode lineshape. If all scattering processes would couple to the same phonon branch, all contributions would be equidistantly spaced in frequency. This is true for outer contributions [Fig. \ref{fig:DecompSplitting}(a)], since the TO splitting along $K-M$ is neglibile. However, the dominant contributions stem from inner processes and therefore, the TO splitting must be taken into account. Since the inner anti-symmetric processes couple to the energetically higher TO branch along $K-\Gamma$, their frequency is upshifted with respect to the center between the symmetric processes. This upshift is a direct measure of the TO splitting and can be easily accessed experimentally. Furthermore, one can also extract the splitting of the electronic bands from the $2D$-mode spectrum, as this parameter is directly connected to the frequency difference between the symmetric processes and the laser-energy dependent shift rate of the $2D$ mode (see SI). Figures \ref{fig:DecompSplitting}(d) and (e) present the measured TO phonon and electronic splitting in comparison with data from DFT+$GW$ calculations. As can be seen, the experimental values are in good agreement with the calculated curve along the inner direction. However, a discrepancy in the TO splitting between theory and experiment can be observed for $\mathbf{q}$ vectors close to $K$. The TO phonon splitting is largest along $\Gamma-K$ and decreases away from this high-symmetry line. Since also phonons away from the high-symmetry direction contribute to the double resonance \cite{PhysRevB.87.075402}, the experimentally measured TO splitting is expected to be smaller than theoretically predicted along $\Gamma-K$. Thus, the theoretical curves should represent lower and upper limits for the experimental values. The fact that the experimental data is also outside those boundaries indicates that the commonly assumed $GW$ correction might still underestimate the TO splitting, which is probably larger than 15\,cm$^{-1}$ close to $K$. Finally, we should note that all results for the $2D$ mode in bilayer graphene are also valid for the $D$ mode.

In conclusion, we demonstrated that the double-resonant $2D$ Raman mode in bilayer graphene is described by three dominant contributions, contradicting all previous works on this topic. We showed that inner processes contribute most to the Raman scattering cross-section, as in single-layer graphene. Moreover, we demonstrated that the TO phonon splitting is of great importance for a correct analysis of the $2D$-mode lineshape. The TO phonon and electronic splitting can be directly extracted from experimental Raman spectra using the presented analysis. Our results highlight the key role of inner processes and finally clarify the origin of the complex $2D$-mode lineshape in bilayer graphene.

FH, PM, and JM acknowledge financial support from the DFG under Grant no. MA 4079/3-1 and the European Research Council (ERC) Grant no. 259286. MC and FM acknowledge support from the Graphene Flagship and from the French state funds (reference ANR-11-IDEX-0004-02, ANR-11-BS04-0019 and ANR-13-IS10-0003-01). Computer facilities were provided by CINES, CCRT and IDRIS (project no. x2014091202).

\bibliographystyle{apsrev4-1}

\newpage
\includepdf{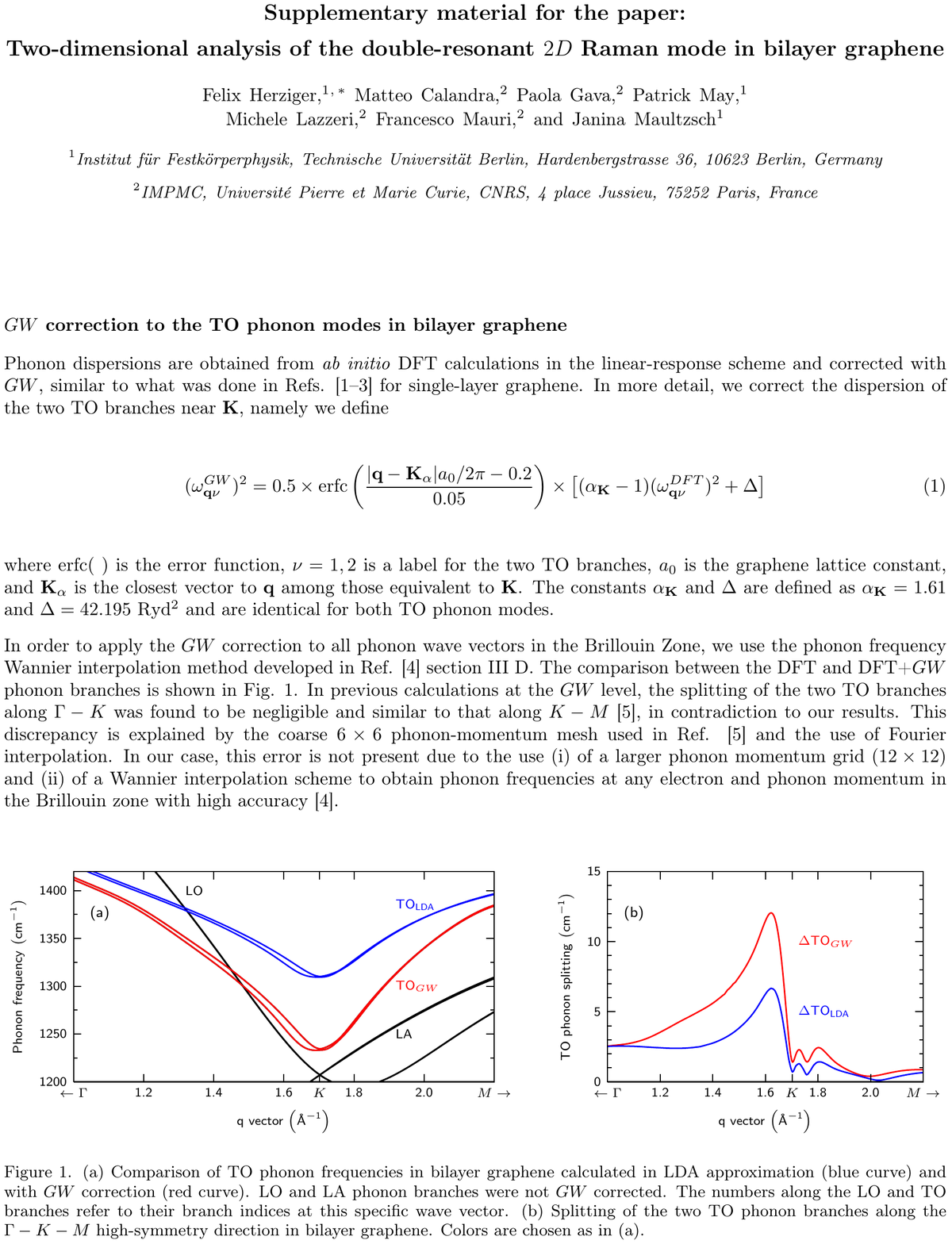}

~
\newpage
\includepdf{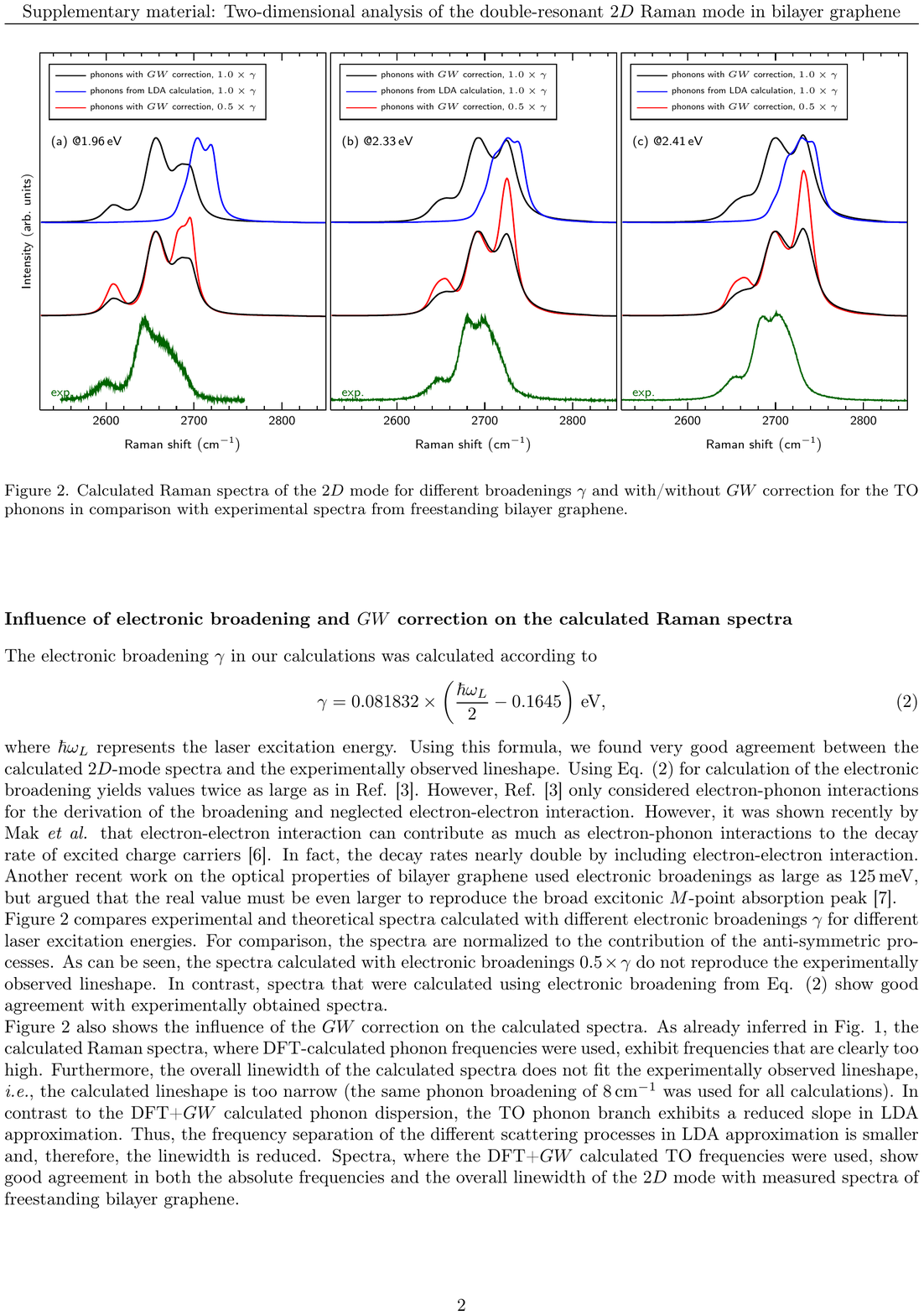}

~
\newpage
\includepdf{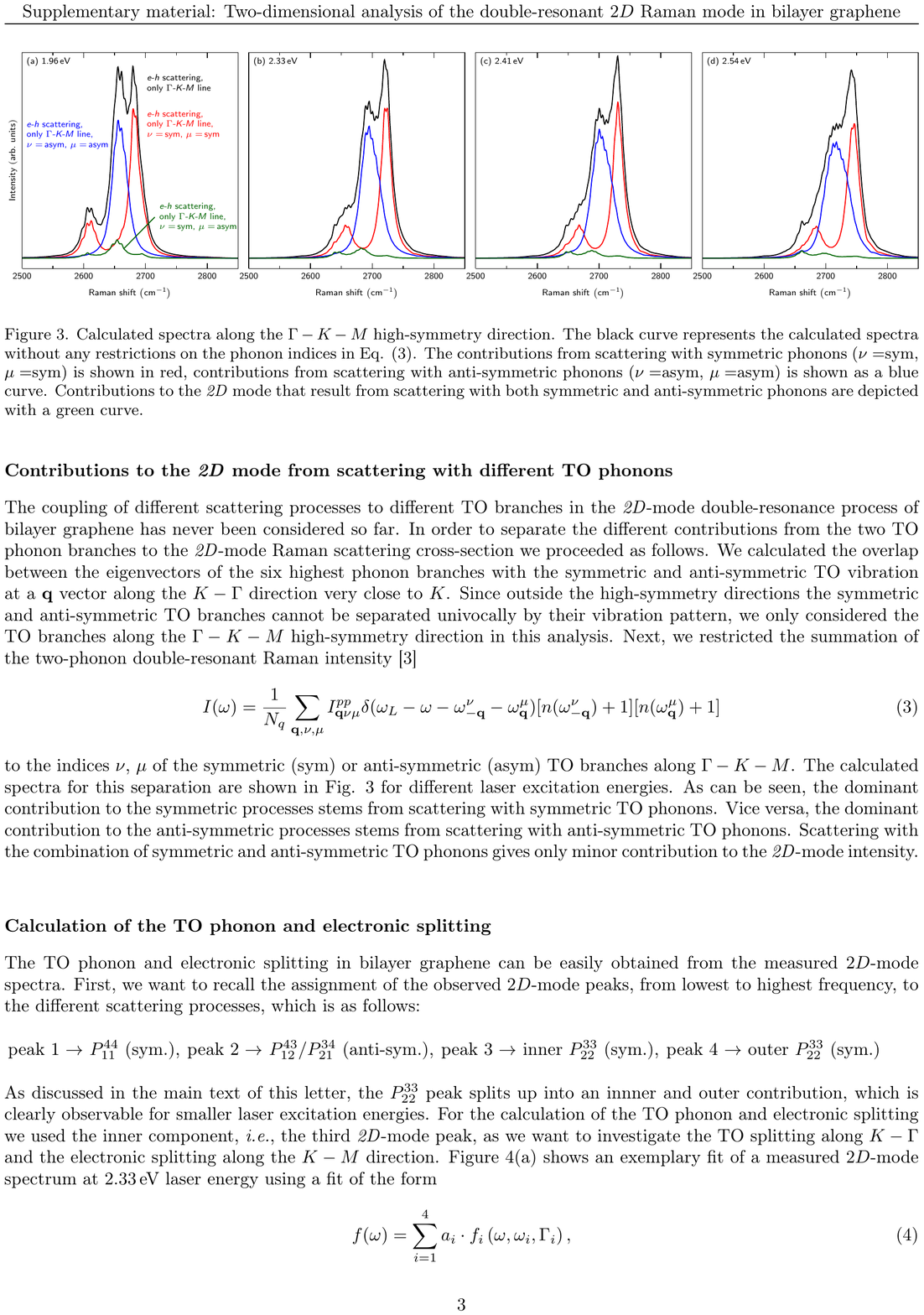}

~
\newpage
\includepdf{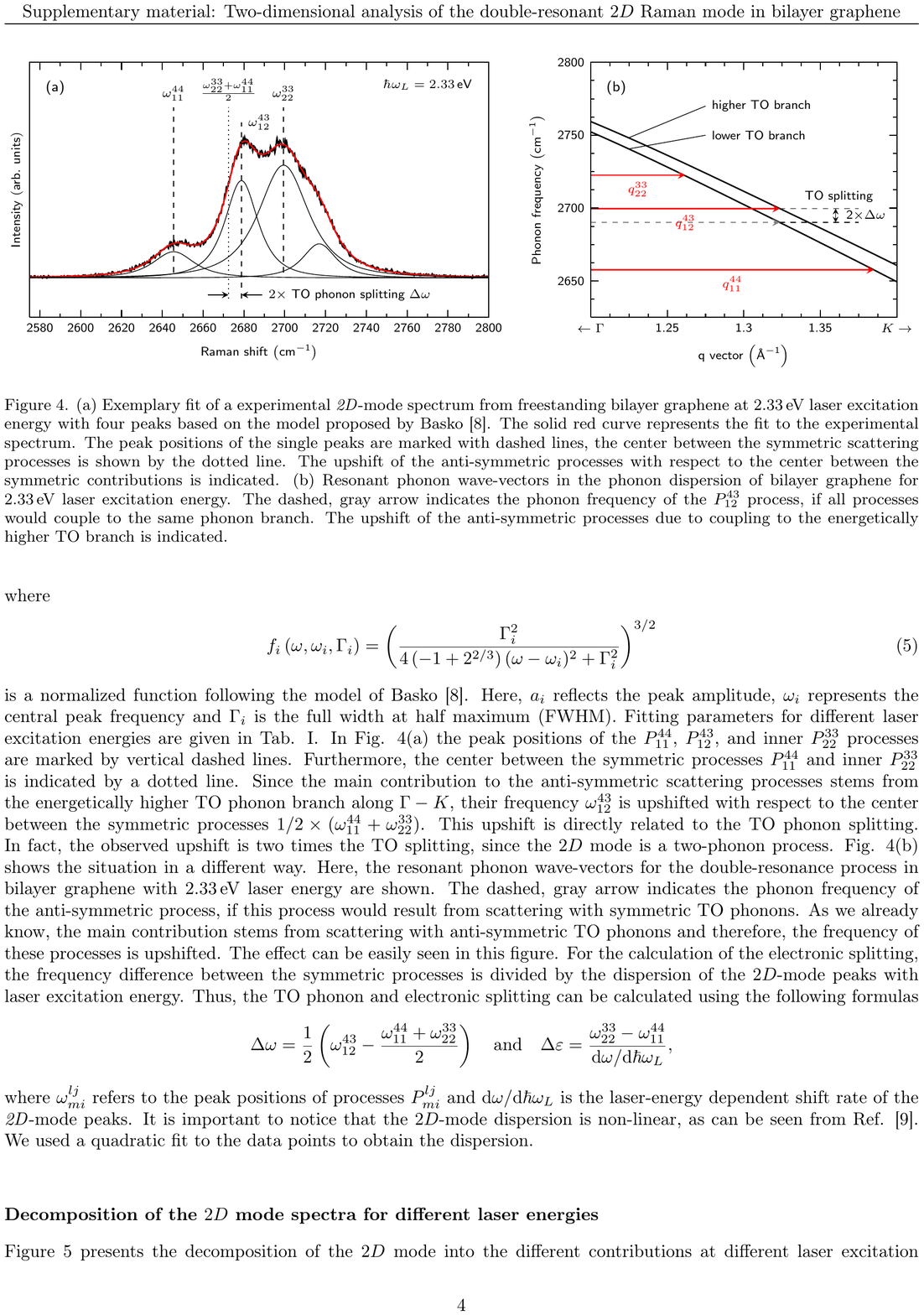}

~
\newpage
\includepdf{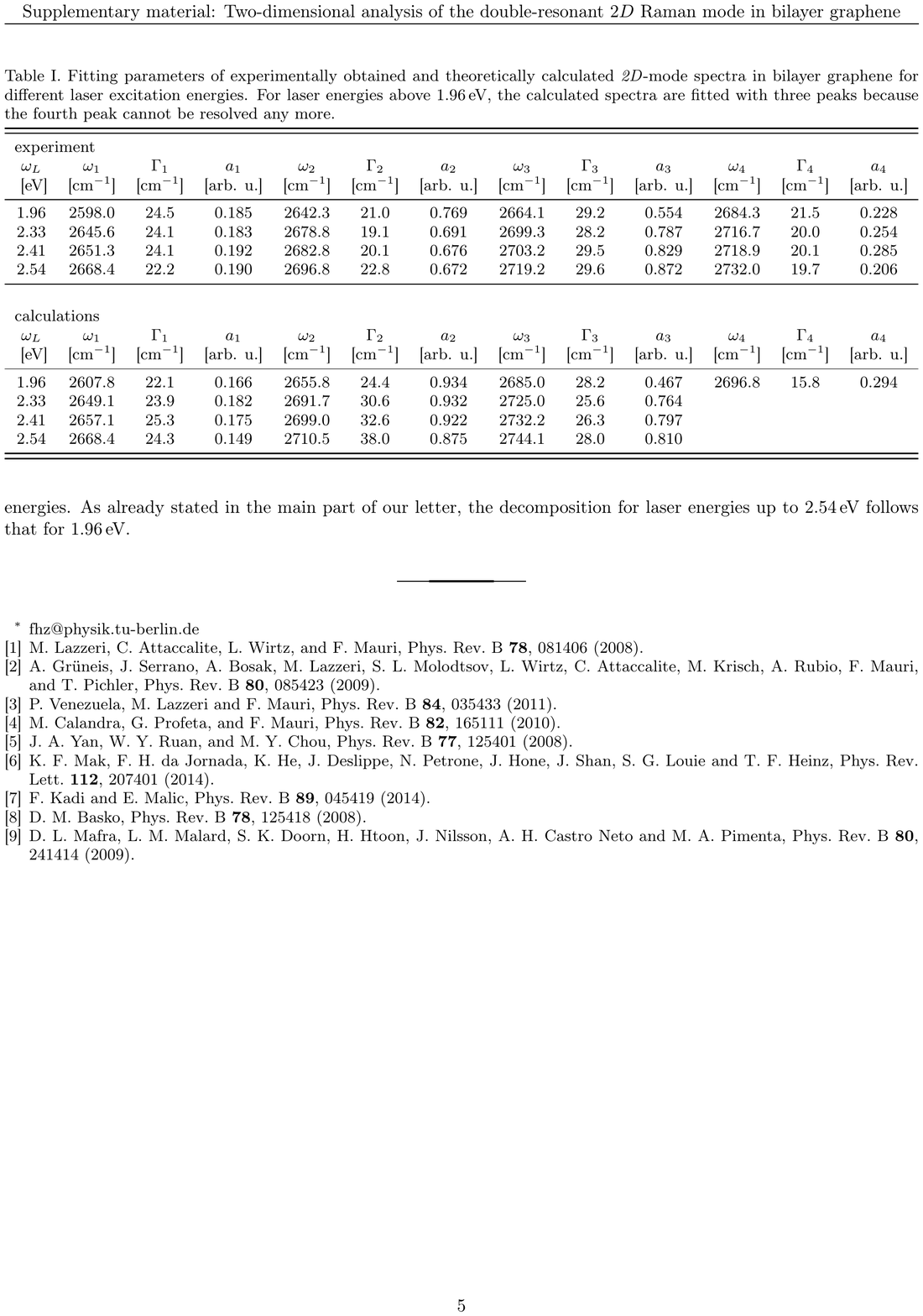}

~
\newpage
\includepdf{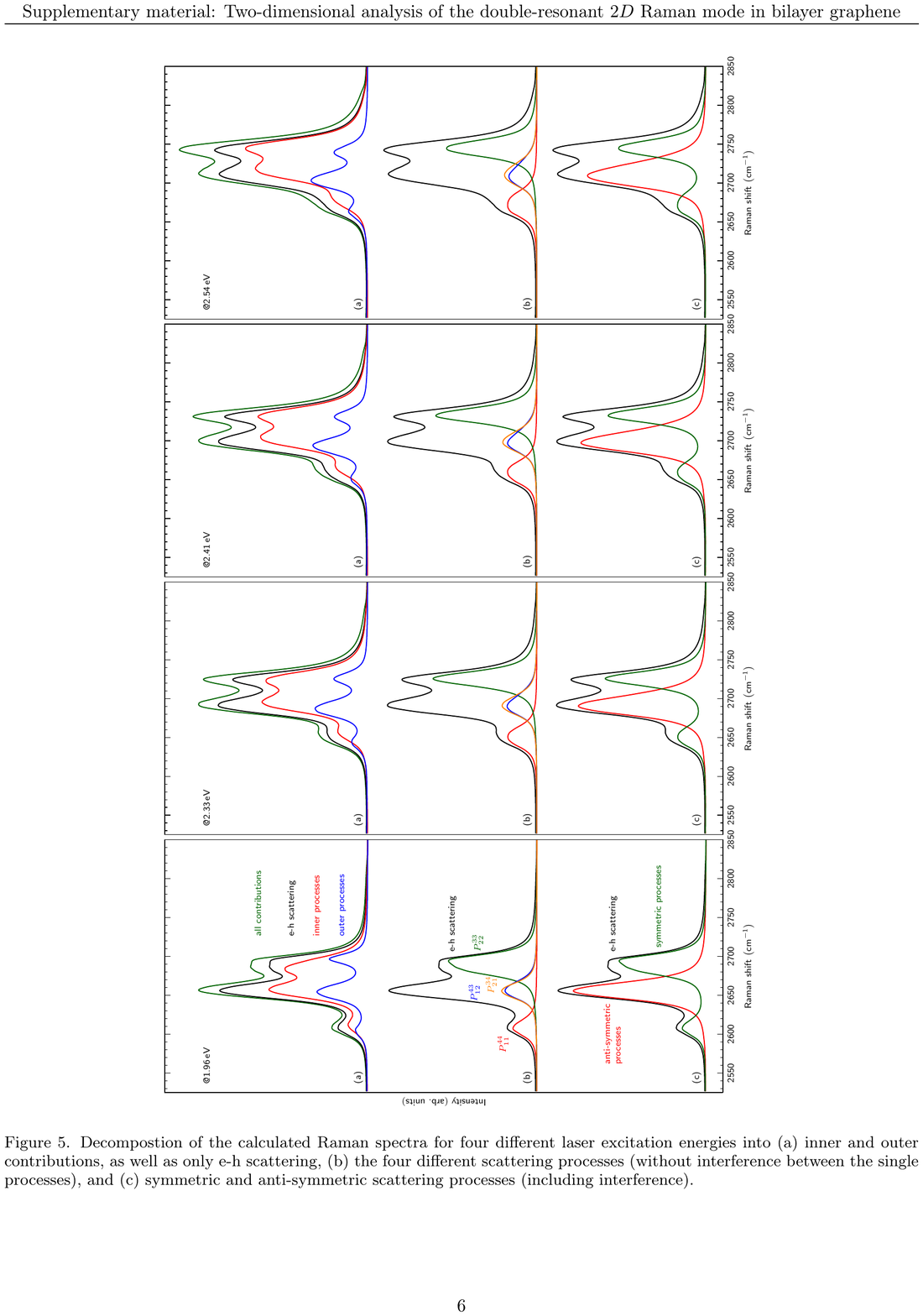}

\end{document}